\documentclass[12pt,a4paper,twoside]{scrartcl}
\usepackage{graphicx}
\usepackage{amsmath}
\usepackage[figuresright]{rotating}
\usepackage{fancyhdr}
\usepackage[width=16cm]{geometry} 
\newcommand{\TheAuthor}{}

\def\skiplinehalf{\medskip\\}

\def\supit#1{\raisebox{0.8ex}{\small\it #1}\hspace{0.05em}}  
\title{Deeply bound meson-nuclear states: theoretical concepts and strategies}
\author{Wolfram Weise
\skiplinehalf
\supit{}Physik-Department, Technische Universit\"at M\"unchen \\
\supit{}D-85747 Garching, Germany}
\date{}
\begin{document}
\maketitle
\begin{abstract}

This presentation reviews recent developments in the understanding of low-energy pion- and kaon-nucleon interactions as they relate to deeply bound pionic atoms and possible antikaon-nuclear bound states. A brief update is given on 1s states of pionic atoms, summarizing their theoretical interpretation based on in-medium chiral effective field theory. This is followed by a state-of-the-art  discussion of low-energy $\bar{K}N$ interactions, with special emphasis on the subthreshold region relevant to the proposed kaon-nuclear systems. 

\end{abstract}
\section{Introduction}
The low-energy interactions of the lightest pseudoscalar mesons (pions and kaons) with
nuclear systems are largely dictated by the spontaneously broken chiral symmetry of QCD.
In the limit of massless up, down and strange quarks, pions and kaons would emerge as massless Goldstone bosons. The small explicit symmetry breaking induced by the $u-$ and $d-$quarks masses, $m_{u,d} < 10$ MeV, moves the pion mass to its empirical (small) value. Likewise, the strange quark mass, $m_s \sim 0.1$ GeV, can still (with caution) be considered small compared to the spontaneous chiral symmetry breaking scale, $\Lambda_\chi = 4\pi f_\pi \sim 1$ GeV, expressed in terms of the pion decay constant $f_\pi \simeq 0.09$ GeV. 

Deeply bound states of pions and kaons in the Coulomb and strong fields of a nucleus represent ideal conditions for investigating the way in which the spontaneous and explicit chiral symmetry breaking pattern of low-energy QCD changes in a nuclear environment. Such systems offer a high-precision testing ground for the detailed behavior of the (energy and momentum dependent) interactions of Goldstone bosons with nucleons in nuclei at varying density. 

Deeply bound pionic atoms have become an important tool for testing chiral pion-nucleus dynamics and the quest for fingerprints of partial chiral symmetry restoration in baryonic matter. The mechanisms at work in forming deeply bound states of pionic atoms are quite different from the one thought to be responsible for the formation of kaon-nuclear bound states. Binding a negatively charged s-wave pion at the surface of a heavy nucleus is a matter of subtle balance between Coulomb attraction and the repulsion resulting from the pion-nuclear strong interaction. In case of a negatively charged (anti)kaon, the driving $\bar{K}$-nucleon interaction is attractive. It is sufficiently strong to generate the $\Lambda(1405)$ just below $\bar{K} N$ threshold. In the region of this resonance, the coupling to the $\pi\Sigma$ channel prohibits the formation of long-lived, narrow $\bar{K}$-nuclear states. However, if the $\bar{K} N$ attraction is still active below the $\pi\Sigma$ threshold, then there is a chance to form narrow $K^-$-nuclear bound states. This is the hypothesis introduced by Akaishi and Yamazaki \cite{AY02}. The present status of data and phenomenology regarding this important issue is reported elsewhere in these Proceedings. The basic question we wish to address here is the following: to what extent does our present knowledge of low-energy $\bar{K} N$ interactions support such expectations? This study is guided by chiral SU(3) effective field theory, representing QCD with strange quarks in the low-energy limit. 

Before turning to this central theme it is useful to give a brief summary of the s-wave pion-nucleus interaction and its detailed energy dependence, driven by the spontaneously broken chiral symmetry of low-energy QCD. This energy dependence is generic to the interaction of Goldstone bosons with matter. The interactions of pions and kaons with nucleons and nuclei are nonetheless quite different in comparison. The reason is the different degree of explicit chiral symmetry breaking, induced by the $u,d$- and $s$-quark masses as they transform into the mass difference between pion and kaon.  

\section{Chiral dynamics
and deeply bound pionic atoms}

The framework here is spontaneously broken chiral symmetry and in-medium chiral perturbation theory, recently extended to finite systems \cite{GRW04}. As in the time-honored Ericson-Ericson approach to pionic atoms, the bound state problem can be written in terms of a Klein-Gordon equation:
\begin{equation}
\left[\vec{\nabla}^2 + (\omega - V_{coul}(\vec{r}\,))^2 - m_\pi^2 - \Pi(\omega - V_{coul}(\vec{r}\,), \vec{r}\,)\right] \Phi(\vec{r}\,) = 0~~~.
\end{equation}
The pion self-energy in the medium, $\Pi(\omega, \vec{r})$, is a function of pion energy $\omega$ and momentum $\vec{q} \rightarrow -i\vec{\nabla}$: 
\begin{equation}
\Pi(\omega,\vec{r}\,) \equiv 2\omega\,U(\omega, \vec{r}\,) = \Pi_S(\omega, \vec{r}\,) + \vec{\nabla}\,\Pi_P(\omega, \vec{r}\,)\cdot \vec{\nabla} + ~~... ~Ê~~.
\end{equation}
The energy-dependent pion-nucleus optical potential $U(\omega, \vec{r}\,) = \Pi(\omega, \vec{r}\,)/2\omega$ is often conveniently introduced to establish contact with an equivalent
Schr\"odinger-type equation. In-medium chiral perturbation theory expresses the pion self-energy as an expansion in powers of two sorts of quantities which are small compared to the chiral scale $4\pi f_\pi$: the pion mass, momentum and energy on one hand, and secondly, the nucleon Fermi momentum \cite{KW01}. The Fermi momentum $p_{F,i}$ for each species of nucleons ($i = p,n$) is related to their densities by $\rho_i = p_{F,i}^3/3\pi^2$. The leading order $"T\rho"$ term as well as double scattering and absorption parts follow
this power-counting pattern. The p-wave self-energy can be taken from \cite{EE66} with the parametrization given in \cite{EW88}. The leading order s-wave self-energy of a $\pi^-$ interacting with a nucleus represented by its proton and neutron density distributions $\rho_{p,n}(\vec{r}\,)$ has the form
\begin{equation}
\Pi_S(\omega, \vec{r}\,) = - {\omega\over 2f_\pi^2}\left(\rho_p(\vec{r}\,) - \rho_n(\vec{r}\,)\right)  + {am_\pi^2 - b\omega^2\over f_\pi^2} \left(\rho_p(\vec{r}\,) + \rho_n(\vec{r}\,)\right) + ~~...~~,
\end{equation}
with the pion decay constant $f_\pi = 92.4$ MeV. The first piece of order $\omega$ is the Tomozawa-Weinberg term which is evidently repulsive for nuclei with a neutron excess. The coefficients of the next-to-leading ${\cal O}(\omega^2, m_\pi^2)$ term involve low-energy constants from pion-nucleon scattering. In NLO chiral perturbation theory:
\begin{equation}
a = 4 c_1 = {\sigma_N \over m_\pi^2} ~~~~, ~~~~ b = 2(c_2 + c_3) - {g_A^2\over 4M_N} ~~~,
\end{equation}
where $\sigma_N \simeq 50$ MeV is the sigma term and $g_A \simeq 1.27$ is the axial vector coupling constant of the nucleon. Empirically one finds $a \simeq b$ from the observed vanishing (within errors)
of the isospin-even pion-nucleon scattering length.

The explicit energy dependence of the s-wave pion-nucleus optical potential reflects the spontaneously broken chiral symmetry and the Goldstone boson nature of the pion. In the exact chiral limit ($m_\pi\rightarrow 0$) and at zero energy, the interaction of the pion with nucleons (and nuclei) vanishes in accordance with Goldstone's theorem. It is important to maintain this basic property in the solution of the bound-state equation (1).

Calculations of deeply bound pionic atoms using the full energy-dependent pion self-energy have been performed in Ref.\cite{KKW03}. Figure 1 shows an example for pionic 1s states in Sn isotopes, for which accurate measurements of the binding energy and width have been performed \cite{Suz04}. The calculation includes a careful assessment of neutron radii through the chain of Sn nuclei. \\

\begin{minipage}[t]{14cm}
\centerline {
\includegraphics[width=6.5cm]{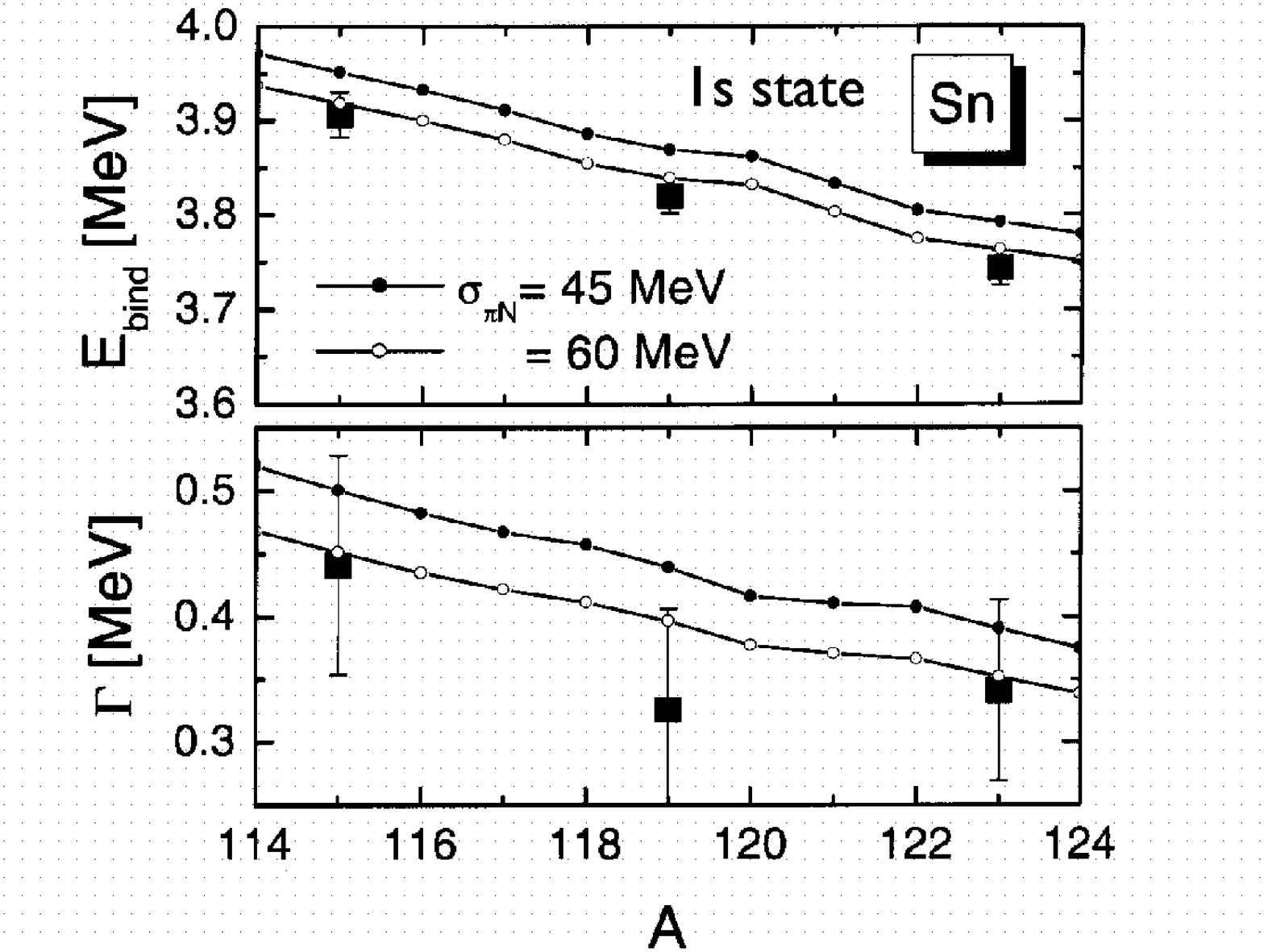}}
\end{minipage}

Figure1: Binding energies and widths of pionic $1s$ states in Sn isotopes. The
curves show predictions \cite{KKW03} based on the explicitly  energy
dependent pionic s-wave polarization operator calculated in two-loop
in-medium chiral perturbation theory \cite{KW01}. The sensitivity to
the $\pi N$ sigma term (input) is also shown. Data are taken from Ref.\cite{Suz04}.\\

The phenomenological analysis of pionic atoms is commonly done not with an explicitly energy dependent 
pion self-energy, but using an energy-independent optical potential with input determined at the 
$\pi N$ threshold. The equivalent, energy-independent (Ericson-Ericson) s-wave potential close to threshold is derived by expanding the $\omega$-dependent self-energy around $\omega = m_\pi$ at zero pion momentum $\vec{q} = 0$. The result deduced from Eq.(3) is
\begin{equation}
\tilde{U}_S(\vec{r}\,) = {\Pi(\omega = m_\pi, \vec{r}\,)\over 2m_\pi\left(1 - {\partial\Pi \over \partial \omega^2}\right)}_{\omega = m_\pi} \simeq {\rho_n - \rho_p\over 4f_\pi^2\left(1 + {b\over f_\pi^2}\rho\right)} \equiv {\rho_n - \rho_p \over 4f_\pi^{*2}(\rho)} ~~~ , 
\end{equation} 
with the nuclear density $\rho = \rho_p + \rho_n$. We make use of $b \simeq a \simeq - \sigma_N/m_\pi^2$. To the order given in Eq.(3), the wave function renormalization factor $\left(1 - {\partial \Pi \over \partial \omega^2}\right)^{-1}$ can then be absorbed in an effective density-dependent pion decay constant \cite{We01},
\begin{equation}
f_\pi \rightarrow f_\pi^*(\rho) \simeq f_\pi\left(1 - 0.2 {\rho\over\rho_0}\right)~~~ .
\end{equation}
A decreasing $f_\pi$ in the nuclear medium (by about 20$\%$ at nuclear matter density, $\rho_0 \simeq 0.17$ fm$^{-3}$) effectively increases the leading repulsive Tomozawa-Weinberg term in the (energy-independent) s-wave optical potential, a welcome feature in view of the longstanding "missing repulsion"
problem that was apparent in the phenomenological analysis of pionic atom data. The $f_\pi \rightarrow f_\pi^*$ effect has been discussed and examined in the context of deeply bound pionic atoms \cite{KY01}. It is also consistent  with new accurate $\pi^{\pm}$-nucleus scattering data taken at very low energies \cite{Fr04}. 

The fact that the characteristic $\omega$ dependence of the s-wave $\pi^-$ self-energy can be translated into an equivalent, energy-independent local potential with a renormalized, density dependent coupling strength, is a consequence of two important features: the Goldstone boson nature of the pion and its nearness to the chiral limit which makes this potential relatively weak and accessible to methods of chiral perturbation theory, and the approximate vanishing of the isospin-even s-wave pion-nucleon scattering length at threshold. For the interactions of (anti-)kaons with nuclei the situation is quite different as we shall now discuss.   

\section{Chiral SU(3) dynamics}

Chiral perturbation theory as a systematic expansion in small momenta and quark masses
is limited to low-energy processes with light quarks. It is an interesting issue to what extent 
its generalisation including strangeness can be made to work. The $\bar{K} N$ channel is of particular interest in this context, as a testing ground for chiral SU(3) symmetry in QCD and for the role of explicit chiral symmetry breaking by the relatively large strange quark mass. However, any perturbative approach breaks down in the vicinity of resonances. 
In the $K^- p$ channel, for example, the existence of the $\Lambda(1405)$ resonance 
just below the $K^- p$ threshold renders SU(3) chiral perturbation theory 
inapplicable. At this point the combination with non-perturbative coupled-channels
techniques has proven useful, by generating the $\Lambda(1405)$ dynamically
as an $I=0$ $\bar{K} N$ quasibound state and as a resonance in the $\pi \Sigma$
channel \cite{KSW95}. Coupled-channels methods combined with chiral $SU(3)$ dynamics have subsequently been applied to a variety of meson-baryon scattering processes with quite some success \cite{KWW97}. The most recent update is given in \cite{BNW05} and reported by B. Borasoy in these
proceedings \cite{Bor05}.

The starting point is the chiral $SU(3) \times SU(3)$ effective Lagrangian. Its leading order terms include the octet of pseudoscalar Goldstone bosons ($\pi, K, \bar{K}, \eta$) and their interactions. Symmetry breaking mass terms introduce the light quark masses $m_u, m_d$ and the mass of the strange quark, $m_s$. The pseudoscalar mesons interact with the baryon octet ($p, n, \Lambda, \Sigma, \Xi$).
through vector and axial vector combinations of their
fields. At this stage the parameters of the theory are the pseudoscalar meson decay constant $f \simeq 90$ MeV and the $SU(3)$ baryon axial vector coupling constants
$D = 0.80\pm 0.01$ and $F = 0.47\pm0.01$ which add up to $D + F = g_A = 1.27$. At next-to-leading order, seven additional constants enter in s-wave channels, three of which are constrained by mass splittings in the baryon octet and the remaining four need to be fixed by comparison with low-energy scattering data.

{\bf Coupled channels.} Meson-baryon scattering amplitudes based on the $SU(3)$ effective Lagrangian involve a set of coupled channels. For example, The $K^- p$ system in the isospin $I=0$ sector couples to the $\pi\Sigma$ channel. 
Consider the $T$ matrix ${\bf T}_{\alpha\beta}(p, p'; E)$ connecting meson-baryon channels $\alpha$ and $\beta$ with four-momenta $p, p'$ in the center-of-mass frame:
\begin{equation}
{\bf T}_{\alpha\beta}(p, p') =\\
 {\bf K}_{\alpha\beta}(p, p') + \sum_\gamma\int{d^4q\over (2\pi)^4} {\bf K}_{\alpha\gamma}(p, q)
 \,{\bf G}_\gamma (q)\,{\bf T}_{\gamma\beta}(q, p')\,\, ,
\label{Teq}
\end{equation}
where ${\bf G}$ is the Green function describing the intermediate meson-baryon loop which is iterated to all orders in the integral equation (\ref{Teq}). (Dimensional regularisation with subtraction constants is used in practise). The driving terms ${\bf K}$ in each channel are constructed from the chiral $SU(3)$ meson-baryon effective Lagrangian in next-to-leading order. In the kaon-nucleon channels, for example, the leading terms have the form
\begin{equation}
{\bf K}_{K^\pm p} = 2{\bf K}_{K^\pm n}  = \mp \,2M_N{\sqrt{s} - M_N\over f^2} + ...\,\, ,
\end{equation}
at zero three-momentum, where $\sqrt{s}$ is the invariant c.m. energy and $f$ is the pseudoscalar meson decay constant. Scattering amplitudes are related to the $T$ matrix by ${\bf f} = {\bf T}/8\pi\sqrt{s}$.
Note that ${\bf K} > 0$ means attraction, as seen for example in the $K^- p \rightarrow K^- p$ channel. Similarly, the coupling from $K^- p$ to $\pi\Sigma$ provides attraction, as well as the diagonal matrix elements in the $\pi\Sigma$ channels. Close to the $\bar{K}N$ threshold, we have ${\bf f}(K^- p \rightarrow K^- p)  \simeq m_K/4\pi f^2$, the analogue of the Tomozawa-Weinberg term in pion-nucleon scattering, but now with the (attractive) strength considerably enhanced by the larger kaon mass $m_K$.
 
One should note that when combining chiral effective field theory with the coupled-channels scheme, the "rigorous" chiral counting in powers of small momenta is abandoned in favor of iterating a subclass of loop diagrams to ${\it all}$ orders. However, the substantial gain in physics compensates for the sacrifice in the chiral book-keeping.  Important non-perturbative effects are now included in the re-summation, and necessary conditions of unitarity are fulfilled. 

{\bf Low-energy kaon-nucleon interactions}. $K^- p$ threshold data have recently been supplemented by new accurate results for the strong interaction shift and width of kaonic hydrogen \cite{Beer04,Ito98}. These data, together with existing information on $K^- p$ scattering, the $\pi\Sigma$ mass 
spectrum and measured $K^- p$ threshold decay ratios, set tight constraints on the theory and have consequently revived the interest in this field. Figure 2 shows selected recent results of an improved calculation which combines driving terms from the next-to-leading order chiral $SU(3)$ meson-baryon Lagrangian with coupled-channel equations \cite{BNW05}. As in previous calculations of such kind, the $\Lambda(1405)$ is generated dynamically, as an $I = 0$ $\bar{K}N$ quasibound state and a resonance in the $\pi\Sigma$ channel. In a quark model picture, this implies that the $\Lambda(1405)$ is not a simple three-quark ($q^3$) state but has a strong $q^4\bar{q}$ component. The detailed threshold behavior of the elastic $K^- p$ amplitude needs to be further examined, in view of the fact that the much improved accuracy of the most recent kaonic hydrogen data from the DEAR experiment indicate an inconsistency with older $K^-p$ scattering data (see ref. \cite{BNW05}). Note that the real part of the $K^-p$ amplitude, when extrapolated into the subthreshold region below the $\Lambda(1405)$, is expected to be large and positive (attractive).  The imaginary part of this amplitude drops at energies below the $\Lambda(1405)$. The dominant $I = 0$ decay into $\pi\Sigma$ is turned off below its threshold at $\sqrt{s} \simeq 1.33$ GeV. The s-wave $K^-n$ subthreshold amplitude is also attractive but less than half as strong as the $K^-p$ amplitude.   \\

\begin{minipage}[t]{7cm}
\centerline {
\includegraphics[width=6.7cm]{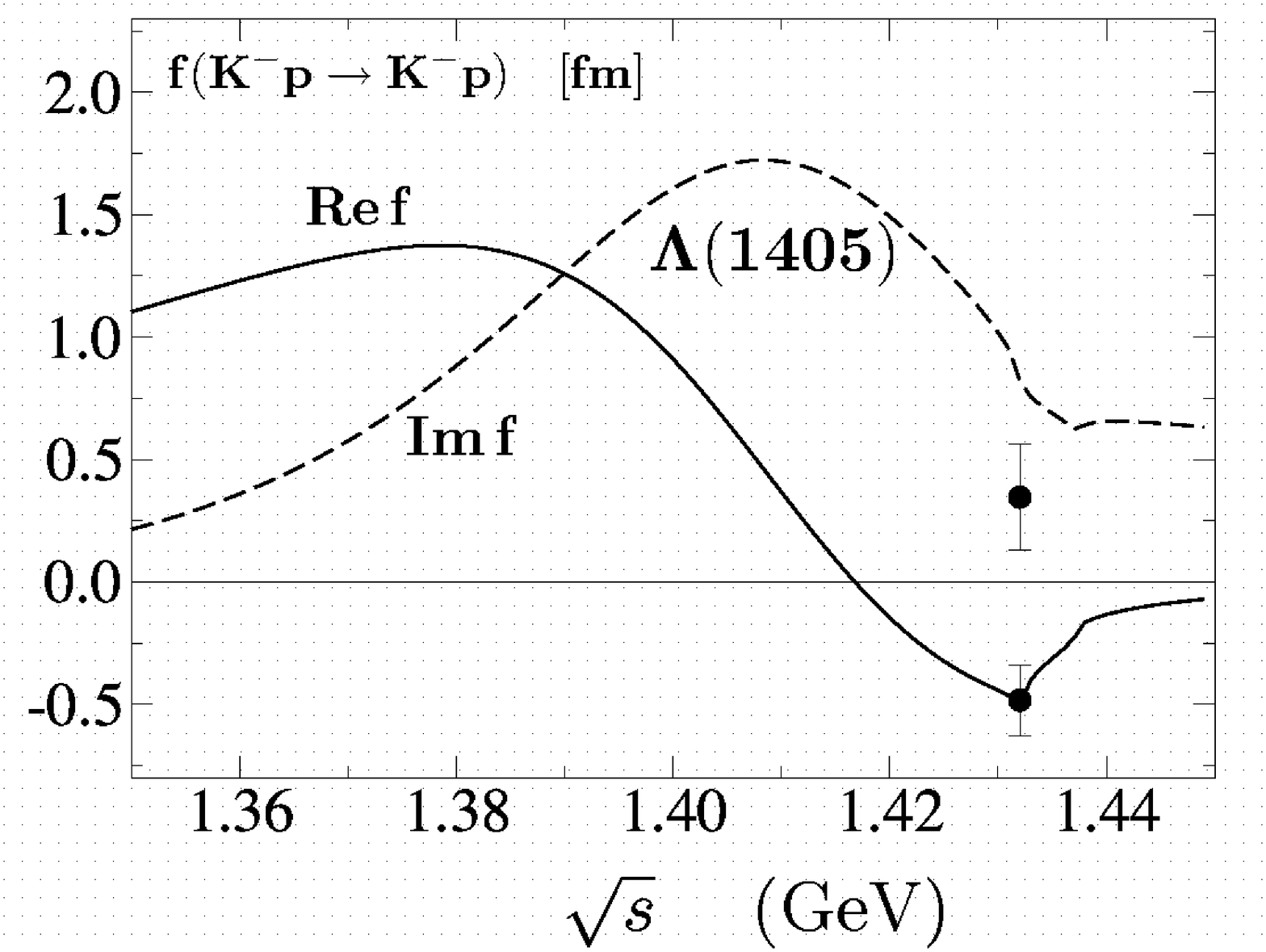}}
\end{minipage}
\hspace{\fill}
\begin{minipage}[t]{8cm}
\centerline {
\includegraphics[width=7cm]{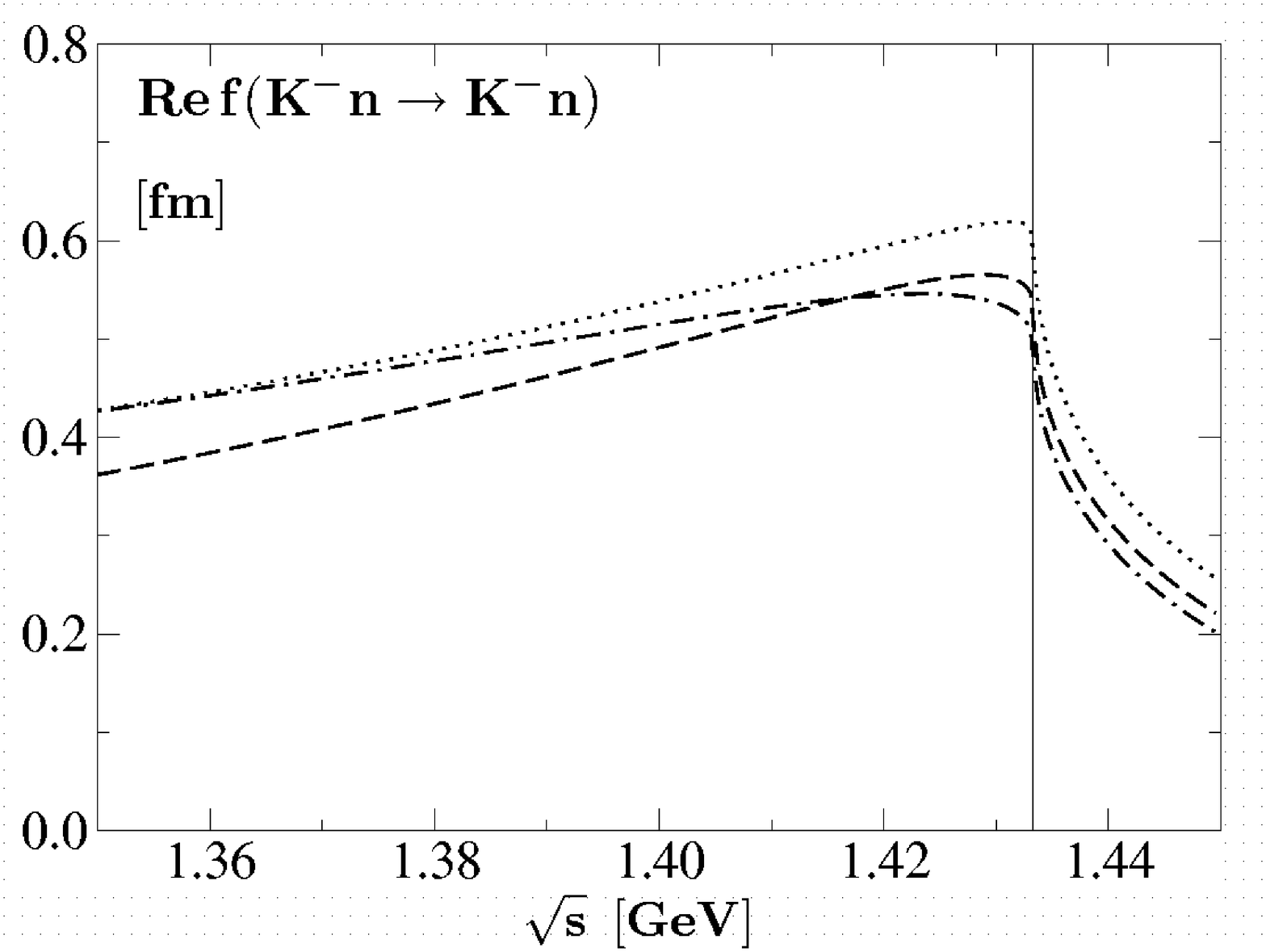}}
\end{minipage}

Figure 2: Left: Real and imaginary parts of the $K^- p$ forward elastic scattering amplitude calculated in the Chiral SU(3) Coupled Channels approach \cite{BNW05}. Real and imaginary parts of the scattering length deduced from the DEAR kaonic hydrogen measurements \cite{Beer04} are also shown. Right: Predicted real part of the $K^-n$ forward elastic scattering amplitude calculated in the same approach \cite{BNW05}. In both figures $\sqrt{s}$ is the invariant $\bar{K}N$ center-of-mass energy.

\section{The $K^-$-nucleus potential and deeply bound kaonic states}

{\bf S-wave potential}. Equiped with the updated s-wave $K^-N$ forward scattering amplitudes and their detailed energy dependence \cite{BNW05}, we can now proceed to estimate the $K^-$nucleus potential derived from the s-wave $K^-$ self-energy in a nuclear medium:
\begin{equation}
\Pi_{K^-} = 2\omega \,U_{K^-} = -4\pi\left[f_{K^-p}\,\rho_p + f_{K^-n}\,\rho_n\right] + \, ...
\end{equation} 
where the additional terms, not shown explicitly, include corrections from Fermi motion, Pauli blocking, two-nucleon correlations etc. The kaon spectrum in matter is determined by
\begin{equation}
\omega^2 - \vec{q}\,^2 - m_K^2 - \Pi_K(\omega,\vec{q}:\rho_{p,n}) = 0.
\end{equation} 
An effective kaon mass in the medium can be introduced by solving this equation at zero momentum: $m_K^*(\rho) = \omega(\vec{q} = 0, \rho)$. Calculations of the spectrum of kaonic modes as a function of density have already a long history. For example, in refs. \cite{WKW96} it was pointed out that, as a consequence of the underlying attractive $\bar{K}N$ forces, the $K^-$ mass at the density of normal nuclear matter ($\rho_0 \simeq 0.17$ fm$^{-3}$) effectively drops to about three quarters of its vacuum value. At this density the $K^-$ in-medium decay width is expected to be strongly reduced because the $K^- N$ energy "at rest" in matter has already fallen below the $\pi\Sigma$ threshold (see Fig. 3). 

\begin{minipage}[t]{14cm}
\centerline {
\includegraphics[width=8cm]{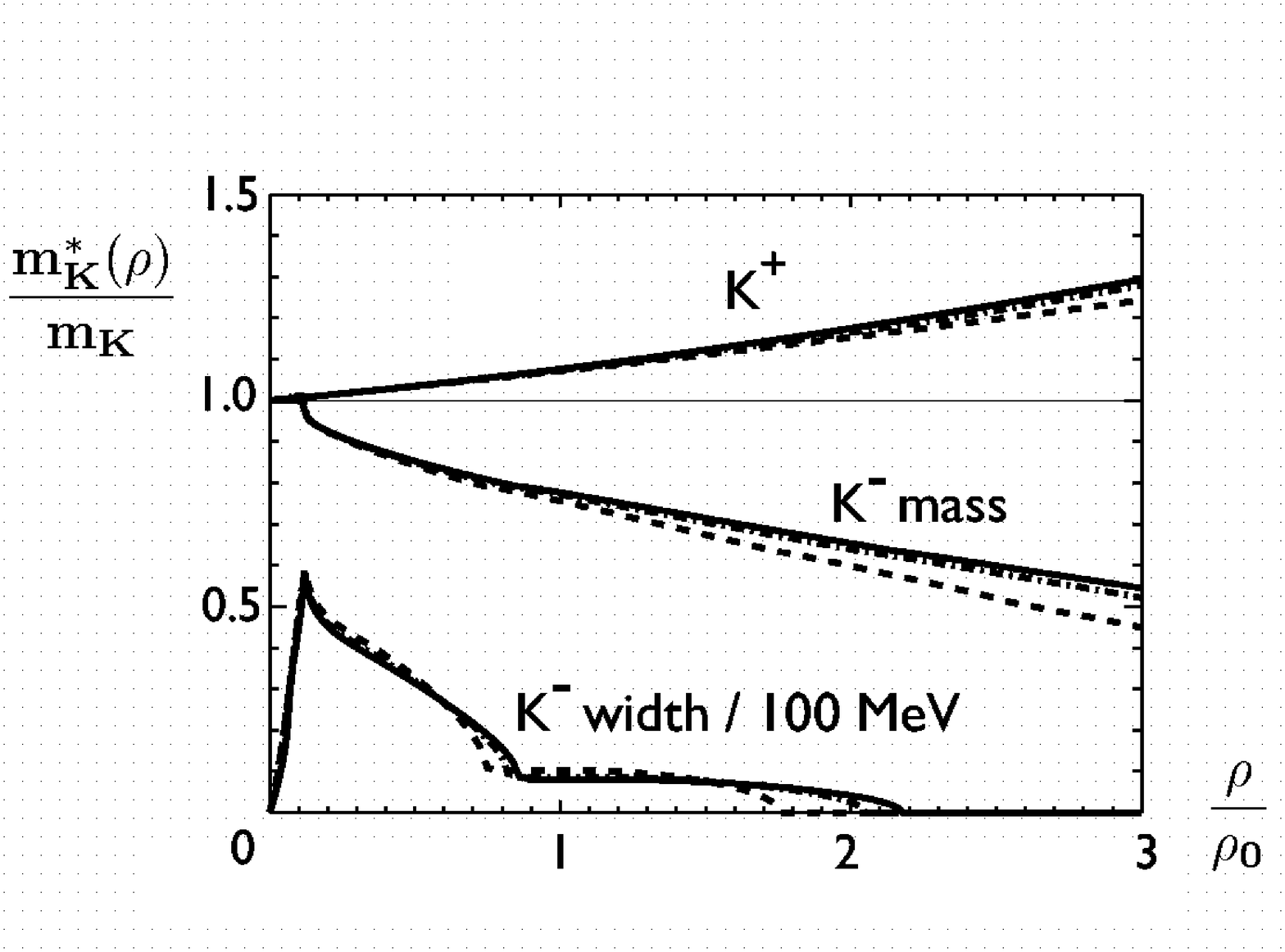}}
\end{minipage}

Figure 3: Calculated in-medium mass $m_K^*(\rho)$ and width of a $K^-$ in symmetric nuclear matter as a function of baryon density $\rho$ in units of nuclear matter density $\rho_0 = 0.17$ fm$^{-3}$. The calculations \cite{WKW96} were performed using in-medium chiral SU(3) dynamics combined with coupled channels and including effects of Pauli blocking, Fermi motion and two-nucleon correlations. Also shown is the in-medium $K^+$ effective mass calculated in the same approach.\\

These calculations do suggest the appearance of a relatively long-lived $K^-$ mode in a dense nuclear medium, with a width reduced to 10 MeV or less once the density reaches that of normal nuclear matter and beyond. The binding potential scales roughly linearly with density.  The predicted potential depth is about -120 MeV at $\rho = \rho_0$. 

Let us now examine the $K^-$-nuclear potential using the new revised analysis \cite{BNW05} of energy dependent subthreshold $K^-p$ and $K^-n$ amplitudes. To leading order in the proton and neutron densities, the local s-wave potential of a $K^-$ interacting with a nuclear core described by local density distributions $\rho_{p,n}(\vec{r}\,)$ is: 
\begin{eqnarray}
U_{K^- core}(\omega, \vec{r}\,) &=& U_p^{(0)}(\omega)\, {\rho_p(\vec{r}\,)\over \rho_0} +   U_n^{(0)}(\omega)\, {\rho_n(\vec{r}\,)\over \rho_0} \nonumber \\
&=&  - {2\pi \over\omega}\left(1 + {\omega\over M_N}\right) \left[ f_{K^-p}(\omega)\,\rho_p(\vec r\,)+  f_{K^-n}(\omega)\,\rho_n(\vec r\,)\right] ~~. 
\end{eqnarray} 
Consider the $K^- N$ amplitudes shown in Fig.2 extrapolated down far below threshold,
e.g. at a kaon bound state energy $\omega \simeq m_K - 100$ MeV. This gives the following estimates:
\begin{eqnarray}
U_p^{(0)}(\omega = m_K - 100 MeV) & \simeq & -(160 + 30\,i)\, MeV~~, \nonumber \\ 
U_n^{(0)}(\omega = m_K - 100 MeV) & \simeq & -(65 + 15\,i)\, MeV~~.
\label{eq:potentials} 
\end{eqnarray} 
Note again the reduced imaginary parts of the potential once the binding is strong enough to move the $K^- N$ energy sufficiently far below the $\Lambda(1404)$ resonance. The imaginary part of the $K^-$-neutron subthreshold amplitude, not shown in Fig.2, turns out to be small, $Im\,f(K^- n) / Re\,f(K^- n) \simeq$ 0.2 at an energy about 100 MeV below $K^- n$ threshold.

The subthreshold amplitudes used in this estimate are constrained \cite{BNW05} by the accurately determined $K^-p$ threshold branching ratios, by the (still poorly known) $\pi\Sigma$ mass spectrum and by the bulk of low energy $K^- p$ scattering data. When implementing the new and precise kaonic hydrogen data from DEAR, the $\pi\Sigma$ mass spectrum experiences a slight downward shift, and consequently increased attraction (by about 20\%) in the potentials (\ref{eq:potentials}). However, the small imaginary part of the $K^- p$ scattering length, resulting from the DEAR analysis and displayed in Fig.2, raises questions of consistency with the previous (low-precision) scattering data, as discussed in detail in Ref.\cite{BNW05}. It is important to clarify this situation with a new generation of even more accurate kaonic hydrogen and deuterium measurements. The constraints imposed by such high-precision data will further reduce uncertainties in the detailed energy dependence of extrapolated subthreshold amplitudes.

{\bf P-wave interactions}. The present estimate (\ref{eq:potentials}), based exclusively on s-wave $\bar{K}N$ amplitudes, would not provide sufficient attraction to explain the strong binding in $K^-$-tribaryon systems reported in refs.\cite{ADY05,Su04}. In comparison with expectations based on the observed $K^-NNN$ signal, this estimate would account for only less than half of the required potential strength. Furthermore, given the fact that the s-wave $K^-n$ (I=1) interaction is substantially weaker than the s-wave interaction in the (dominantly I=0) $K^-p$ channel, one would expect weaker binding in the $K^-pnn$ than in the $K^-ppn$ system, with an estimated ratio $U(K^-ppn)/U(K^-pnn)\sim 4/3$. This is opposite to what seems to be observed \cite{ADY05, Su04}.

At this point the relevance of p-wave interactions and the role of the $\Sigma^*(1385)$ has been emphasized by Wycech and Green in Ref.\cite{GW05}. If the $K^-$ binding is sufficiently strong so that the $\bar{K}N$ energy falls below the $\Sigma^*$ resonance, then the p-wave $K^-N$ interaction would  turn from weakly repulsive at threshold to strongly attractive below the $\Sigma^*$, a situation reminiscent of the $\Delta(1230)$ in the pion-nucleon interaction. The $\Sigma^*$-resonant p-wave gives stronger weight to $K^-n$ than to $K^-p$ channels. Its $\vec{\nabla}\rho(\vec{r}\,)\cdot\vec{\nabla}$ structure enhances the effect at the surface of compressed nuclei with large density gradients. Using SU(3) to estimate the $KN\Sigma^*(1385)$ coupling strength relative to that of the $\pi\Lambda\Sigma$ vertex,  Wycech and Green find indeed that this p-wave mechanism provides an additional source of attraction which can potentially account for the stronger binding in $K^-pnn$ as compared to $K^-ppn$.    These considerations call for an improved treatment of non-local, energy-dependent interactions in order to promote our understanding of deeply bound $K^-$- few nucleon systems, should their existence be confirmed. 

{\bf Outlooks}. In view of the strongly attractive $\bar{K}N$ subthreshold interaction, it is suggestive \cite{AY02,ADY05} to look for even stronger attraction in nuclear clusters with more than one $\bar{K}$, such as $K^-K^-NN$. This raises the question about processes by which several kaons can be produced under kinematical conditions which minimise their momentum relative to a nuclear target.
An example of such a reaction is antiproton-proton annihilation into two kaon-antikaon pairs:
\begin{equation}
\bar{p}p \rightarrow K^-K^-K^+K^+ \,\, .
\end{equation}
The threshold energy of the antiproton in the lab frame for this process is low, $T_{lab}(\bar{p})\simeq 204$ MeV. A rough estimate for the cross section close to threshold. based on scarce data \cite{Comp84} and a model of $\bar{p}p$ annihilation mechanisms \cite{MVW91} gives
\begin{equation}
\sigma(\bar{p}\,p \rightarrow 2K^+\,2K^-) \sim 10 \,\mu b
\end{equation}
by order of magnitude. While this is only a small fraction ($\sim 10^{-4}$) of the total $\bar{p}p$ annihilation cross section, the reaction can be uniquely identified by measuring the two produced $K^+$'s.

Now consider the same reaction with a nuclear target and assume that there is strong clustering of the two produced $K^-$'s with two nucleons in the nucleus. For example, if the formation of
a $K^-K^-pp$ cluster releases several hundred MeV of binding energy, while at the same time the two $K^+$'s experience only weak repulsion in the nucleus as indicated in Fig.3, pronounced signals for events of the kind
\begin{equation}
\bar{p}A \rightarrow K^+K^+ + [K^-K^-pp(A-3)]
\end{equation}
should be observable in the spectrum of the two $K^+$'s detected in coincidence.  Even with a stopped initial antiproton, the mere appearance of two $K^+$'s in the final state would already indicate strong binding of a pair of $K^-$'s in the target nucleus.  The kinematical conditions provided with low-energy antiprotons are evidently suitable for keeping momentum transfers minimal among the participants. It would be interesting to study the feasibilty of such a measurement at the future antiproton facility at FAIR / GSI.\\

\vspace{0.8cm}

\noindent
{\bf Acknowledgements}\\
Instructive discussions with Y. Akaishi, P. Kienle and T. Yamazaki are gratefully acknowledged. Special thanks go to Bugra Borasoy and Robin Ni{\ss}ler for their cooperation.

\end{document}